\documentclass[12pt,preprint]{aastex}

\usepackage{emulateapj5,graphics} 

\newcommand{\ks}{K_{\rm s}}

\slugcomment{Draft of \today}

\shorttitle{Granular Extinction in the Galactic Centre}
\shortauthors{Gosling, Blundell \& Bandyopadhyay}

\begin{document}

\title{Complex small-scale structure in the infrared extinction
towards the Galactic Centre}

\author{Andrew J.\ Gosling\altaffilmark{1}, Katherine M.\
  Blundell\altaffilmark{1} and  Reba Bandyopadhyay\altaffilmark{1}}

\altaffiltext{1}{University of Oxford, Department of Physics, Keble
  Road, Oxford, OX1 3RH, U.K.}

\begin{abstract}

A high level of complex structure, or ``granularity'', has been
observed in the distribution of infrared-obscuring material towards
the Galactic Centre (GC), with a characteristic scale of $5\arcsec -
15\arcsec$, corresponding to $0.2 - 0.6\,{\rm pc}$ at a GC distance of
$8.5\,{\rm kpc}$. This structure has been observed in ISAAC images
which have a resolution of $\sim0.6\arcsec$, significantly higher than
that of previous studies of the GC.

We have discovered granularity throughout the GC survey region, which
covers an area of $1.6^{\circ} \times 0.8^{\circ}$ in longitude and
latitude respectively ($300\,{\rm pc} \times 120\,{\rm pc}$ at
$8.5\,{\rm kpc}$) centred on Sgr A*.  This granularity is variable
over the whole region, with some areas exhibiting highly structured
extinction in one or more wavebands and other areas displaying no
structure and a uniform stellar distribution in all wavebands.  The
granularity does not appear to correspond to longitude, latitude or
radial distance from Sgr A*. We find that regions exhibiting high
granularity are strongly associated with high stellar reddening.
\end{abstract}



\section{Introduction}
\label{sec:intro}

Observations towards the Galactic Centre (GC) are extremely difficult
because of high levels of extinction from intervening material in the
Galactic Plane. The gas and dust which form the inter-stellar medium
(ISM), if distributed homogeneously along our line of sight, would
produce $\sim$ 100 magnitudes in visual extinction \citep{laun02}, in
stark contradiction to observations \citep[e.g.][]{riek85, catc90,
blum96, dutr03} which indicate the true average extinction is $A_V
\leq 30$. The explanation inferred for this, supported by numerous
observations at different wavelengths \citep{catc90, lis94, dutr03} is
that the ISM towards the GC is distributed non-uniformly.  Hitherto,
no extinction map towards the GC has had an angular resolution $\leq
1\arcmin$.\\

We describe the results derived from new, high-resolution
near-IR images of 26 fields within the nuclear bulge. These VLT-ISAAC
fields \citep{band05}, are $2.5 \times 2.5$ arcmin$^2$ in size, with a
plate scale of $0.1484\arcsec$ per pixel. The fields are distributed
throughout an area $1.6^{\circ} \times 0.8^{\circ}$ $(l,b)$ centred on
Sgr A*, avoiding areas of known high star formation. We obtained a
limiting magnitude of $J=23$ $(S/N=5)$, $H=21$ and $\ks =20$
$(S/N=10)$, taken on nights with seeing $\le0.6\arcsec$.


The purpose of this Letter is to report the discovery and quantify
the size scales of structure in the dust distribution unresolvable in
previous studies. A follow-up paper (Gosling et al. in prep.)
presents extinction maps and a detailed analysis of comparisons with
previous, lower-resolution maps such as that of 2MASS and DENIS.


\section{Granularity}
\label{sec:granularity}

\begin{figure*}[t]
\begin{center}
\scalebox{0.80}{\includegraphics{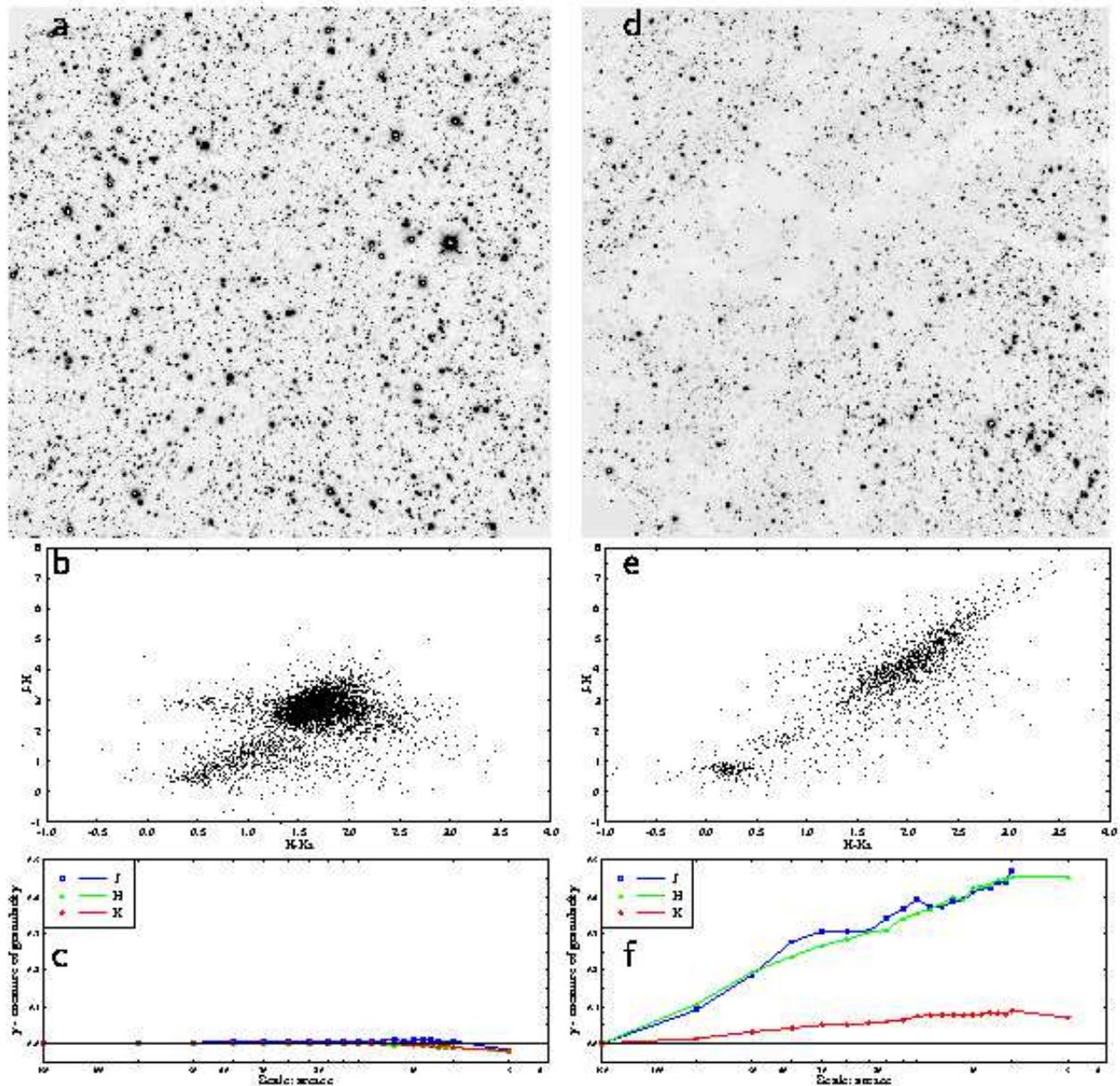}}
\caption{a) $\ks$ image of a field with no apparent structure in the
stellar distribution. b) C-C diagram of the stars in the field shown
in (a). There are two main loci of stars: the local population to the
bottom-left and the GC population in the centre. c) The measure of
granularity for the field shown in (a). In all three bands it does not
deviate from zero, as expected for a random distribution. d) $\ks$
image of a field with obvious regions of low stellar density compared
to the field average. e) C-C diagram for the field shown in (d). Note,
compared to the C-C diagram in (b), the locus for the GC stars is
extended to high reddening. f) The measure of granularity for the
field shown in (d) shows that there is measurable structure in all
three bands.}
\label{plotone}
\end{center}
\end{figure*}

Examination of the VLT-ISAAC fields of the GC reveals that the degree
of reddening seen in colour-colour (C-C) and colour-magnitude (C-M) diagrams of
the field populations varies widely (see Fig.\,\ref{plotone}b and
e). This property is not evident in all fields.  C-C plots
for some of the fields (such as Fig.\,\ref{plotone}b) show two loci of
stars, one for foreground stars, and another with reddening consistent
with traditional values for the GC.  C-C plots of other
fields (such as Fig.\,\ref{plotone}e) show the locus of the non-local
stellar population to be greatly extended along the reddening vector,
and containing stars with reddening considerably in excess of the
traditional GC reddening values.


Visual inspection shows that fields where the reddening appears to
vary substantially within the field also display structure in the
stellar distribution on sub-arcmin scales. This structure takes the
form of filaments and clumpy regions within which the stellar density
is dramatically reduced compared with the field average.  We present
evidence in \S\ref{sec:granred} that these structures are related to
the increased and varying reddening in these fields. The fine scale of
the structures we have discovered were inaccessible to previous
lower-resolution surveys such as 2MASS.


To quantify ``granularity'', we used statistical analyses to
measure the structure of the observed stellar distribution as a
function of wavelength.  \cite{wall03} present techniques for
statistical analysis of 2-dimensional distributions of points. Their
method is used to detect structure in galaxy distributions over the
whole sky.  To apply this test it is assumed that the underlying
stellar distribution is random, and that the cause of the structure
arises from the intervening extinction.
The $\ks$-band images will be less affected by extinction than the
$J$- and $H$-bands. If the underlying distribution of stars is random,
then the $\ks$ images should show a measure closer to random than
equivalent images in $J$ and $H$.\\

The test divides each field into a grid of cells, whose sizes were
systematically varied from $150\arcsec$ to $1\arcsec$ to quantify the
size scale of the structures.  The mean star density per square arcsec
was determined by dividing the number of stars in the entire field by
the field area. The deviation of the number of stars observed in a
grid cell was compared to the mean number per cell. The deviations
from the mean were summed and the variance $\mu_2$ of all of the cells
was calculated.  Because the mean number of stars per unit area is a
constant, and it is the size of a cell that is altering, the variance
is normalised to the size of the grid cell, as expressed in the
formula below,

\begin{equation}
y = \frac{\mu_2 - \bar{n}}{\bar{n}^2} 
\end{equation}
where $y$ is the measure of the deviation from randomness for the whole field,
i.e.\ granularity, and $\bar{n}$ is the expected number of stars in one grid
cell. \\

\begin{figure*}[t]
\begin{center}
\scalebox{0.50}{\includegraphics{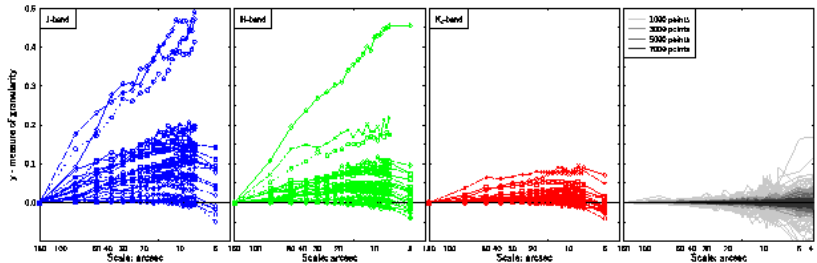}}
\caption{Statistical measures of the granularity $y$, in the GC
fields.  From left to right, these are for $J$, $H$, $\ks$-band
observations.  The rightmost panel is for test fields as described in
the text; the darker colours in this panel represent the test fields
with higher stellar densities.  The granularity measured for our GC
fields is truncated below 3-$\sigma$ of the average stellar
separation.  }
\label{plottwo}
\end{center}
\end{figure*}

The number of stars ranges from 1211 to 5592 in $J$ images, 1792 to
6515 in $H$ and 3005 to 9077 in $\ks$.  To act as comparison
data-sets, we produced 400 test fields with an equivalent spatial
distribution to the observed fields using a computer random number
generator to position points in a grid, 100 each with 1000, 3000, 5000
and 7000 points. We applied the same statistical test to these
simulated fields as to the GC fields.  Comparison of the simulated and
actual data allowed us to determine whether the degree of granularity
measured in a GC field, $y$, was simply a result of low stellar
density and as such a statistical effect, or the result of a
non-random apparent stellar distribution. The maximum values of $y$
for the random fields were 1.8, 0.08, 0.05 and 0.04 for the fields
with 1000, 3000, 5000 and 7000 points respectively, at a scale of
$2\arcsec$ (Fig. \ref{plottwo}).


\section{Results}
\label{sec:analysis}

\subsection{Uniform Fields}

Of our 26 GC fields, 5 showed no measurable difference in granularity
to the simulated random fields (Fig.\,\ref{plotone} (left)). The
stellar distribution of these fields is indistinguishable from random
in all three wavebands indicating that the extincting material has no
structure on scales measurable at the spatial resolution of our data.
The C-C and C-M diagrams of these 5 fields indicate that there is a
single value of reddening for these stars (consistent with an
approximate value of $A_K = 2.5$ for the GC).  This single value for
the reddening agrees with the observed lack of granularity across
these fields.

\subsection{Granular Fields}
The remaining 21 fields all exhibit some degree of structure in
their stellar distribution and show a greater range of reddening
in their stellar population than fields without granularity.

Ten fields show only a small level of granularity, with a
maximum $y$ value of 0.1 in $J$, reducing in $H$ and
showing zero granularity in $\ks$. The C-C and
C-M diagrams for these ten fields show reddening that is
slightly increased and more varied than the fields with no
granularity (\S\,3.1) ranging over $\sim1$ magnitude in $\ks$.  

The remaining 11 fields show granularity in all three bands. The
granularity is highest in $J$, intermediate in $H$ and lowest in
$\ks$; as expected the longer wavelength observations  are
less affected. Thus the $\ks$-band images will have an apparent
measured stellar distribution closer to the true underlying
distribution. The C-C and C-M diagrams for
these fields show a wide range of values of reddening within
individual fields, with the values of reddening 2--3$\times$ that
expected for the GC (see Fig. \ref{plotone}e).

\subsection{Granularity Properties}
\label{sec:granred}
\begin{figure*}
\begin{minipage}{\columnwidth}
\includegraphics[width=\columnwidth]{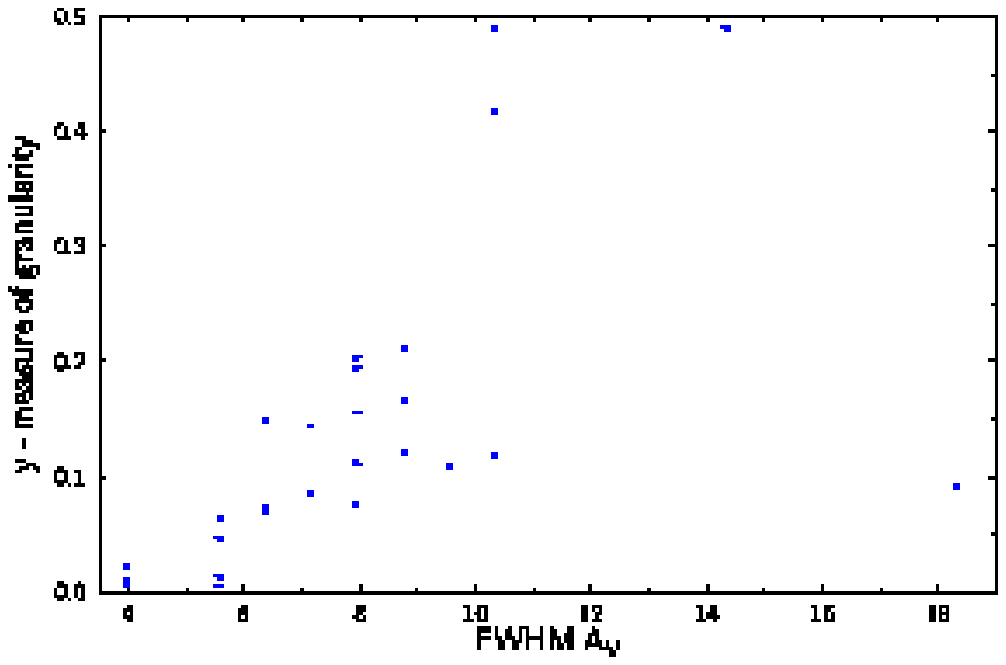}
\caption{Relationship between level of granularity $(y)$ in a field
  and the dispersion in the extinction distribution measured in the
  same field.}
\label{plotthree}
\end{minipage}
\end{figure*}

The nature of the {\em wavelength dependence of the granularity},
namely highest $y$-values in $J$, intermediate in $H$ and low/absent
in $\ks$, indicates it is an effect of intervening extincting
material.  Fig\,\ref{plotone} demonstrates this relationship between
extinction and the granularity in the stellar distribution.  Further,
the association of high and varied reddening in C-M and C-C diagrams
with significant measured granularity (i.e.\ high value of $y$)
provides independent evidence for the relationship between extinction
and granularity. These data indicate that the cause of both observed
effects is the same; namely, that both are a result of dust and gas
structures in the GC.  In Fig\,\ref{plotthree} we plot the range of
the FWHM of $A_K$ for each field versus their measured granularity,
these values are given in Table\,1 for all fields.  The
extinction values were calculated using the near infrared colour
excess (NICE) method of \citet{lada94}, with model giant branch
colours supplied by P. Podsiadlowski ({\it priv comm}) for
comparison. The extinction law of \citet{riek85} was used to calculate
extinction values from the colour excesses.  A detailed spatial
analysis of this relationship, including comparison to reference giant
branches and analysis of the extinction law towards the GC, is
presented in the follow up paper (Gosling et al.)  together with
extinction maps of the individual fields.


\begin{table*}
\begin{tabular}{rccrrc}
\tableline
Field & RA & Dec & Modal & $\Delta\,A_K$ & $y$  \\
name & (J2000) & (J2000) & $A_K$ & & \\
\tableline
  2 & 266.76 & -28.88 &       & 0.892 & 0.156 \\
 25 & 265.96 & -29.16 & 2.765 & 0.446 & 0.011 \\
 35 & 266.05 & -29.45 &       & 0.981 & 0.212 \\
 56 & 266.13 & -29.31 &       & 1.605 & 0.490 \\
 58 & 266.19 & -29.22 &       & 0.892 & 0.194 \\
 72 & 266.33 & -29.09 &       & 0.981 & 0.167 \\
 83 & 266.11 & -29.65 &       & 0.981 & 0.120 \\
 84 & 266.78 & -28.82 &       & 1.070 & 0.108 \\
 89 & 266.69 & -28.28 &       & 0.624 & 0.046 \\
 95 & 266.66 & -28.93 &       & 0.713 & 0.074 \\
130 & 266.52 & -29.11 &       & 0.713 & 0.070 \\
137 & 265.78 & -29.40 &       & 0.892 & 0.203 \\
151 & 266.24 & -29.09 &       & 0.892 & 0.112 \\
162 & 266.48 & -28.34 & 2.052 & 0.624 & 0.005 \\
174 & 267.13 & -28.38 &       & 2.051 & 0.091 \\
183 & 266.23 & -29.32 &       & 0.892 & 0.076 \\
195 & 266.41 & -28.67 & 2.141 & 0.624 & 0.013 \\
224 & 266.12 & -28.94 &       & 1.160 & 0.417 \\
243 & 266.10 & -29.91 & 2.409 & 0.446 & 0.007 \\
289 & 266.76 & -28.24 &       & 0.802 & 0.086 \\
312 & 266.39 & -29.21 &       & 1.159 & 0.119 \\
339 & 266.14 & -28.74 &       & 0.713 & 0.148 \\
391 & 265.55 & -29.61 & 1.873 & 0.446 & 0.021 \\
394 & 266.03 & -29.45 &       & 1.159 & 0.490 \\
423 & 266.30 & -29.28 &       & 0.624 & 0.064 \\
486 & 266.25 & -29.00 &       & 0.802 & 0.144 \\
\tableline
\end{tabular}
\hsize\columnwidth
\noindent \small {$\Delta\,A_K$ is the FWHM of the extinction distribution (excluding the foreground
x  population). $y$ is the maximum measured in the field. Values of
  modal $A_K$ are only quoted for fields with negligible granularity
  for which an average $A_K$ is meaningful.} \\
\label{table1}
\end{table*}


The characteristic scale for the size of the extincting regions is $
5\arcsec - 15\arcsec$ (Fig. \ref{plottwo}), the same scale being
measured in all three bands in all fields.  This angular size
corresponds to a physical scale of $0.2\,{\rm pc} - 0.6\,{\rm pc}$ at
a GC distance of $8.5\,{\rm kpc}$. The average separation of stars
over all fields is $2.72\arcsec$ in $J$, $2.29\arcsec$ in $H$ and
$1.94\arcsec$ in $\ks$. Fig. \ref{plottwo} shows that we can exclude
the possibility that the observed granularity is due to the intrinsic
stellar distribution at the 99.7\% confidence level in all three
bands.  Note that the $y$ parameter in Fig. \ref{plottwo} is truncated
below the $3\sigma$ value of the average stellar separation.

Consideration of the granularity measure in all the fields with their
positions on the sky showed that the granularity does not trace a
field's position on the sky with respect to Sgr A*; no trend is seen
with Galactic $l$, $b$, or projected radial distance
from Sgr A*.  In addition, there appears to be no relation between the
stellar density of a field and its position with respect to Sgr A* or its
level of granularity.

We investigated whether there was any clear link between the
granularity and the extincting material observed in surveys at other
wavebands. Comparison to ISOGAL MIR surveys \citep{omon03et, vanl03}
did not show evidence that the granularity traced areas of
strong MIR emission. A large number of structures observed in our
study seemed to be narrow and elongated in appearance, so a comparison
of the levels of granularity was made with non-thermal filaments and
other features also observable in the radio \citep{vall04, yuse04,
mezg96}. No correspondence of radio structure with levels of
granularity was observed.  The resolution of published IR
extinction studies has been insufficient to allow comparison to the
structures we have discovered in our high-resolution ($< 0.6\arcsec$)
survey.

\section{Conclusions}

Complex structure in the extincting material towards the Galactic
Centre has been observed and measured for the first time on arcsec
scales. The angular scale of the structures is $5\arcsec - 15\arcsec$
which corresponds to a physical scale of $0.2\,{\rm pc} - 0.6\,{\rm
pc}$ at a GC distance of $8.5\,{\rm kpc}$. 

Granularity is only apparent in fields in which there is high and
variable reddening, (derived from the observed stellar C-C
and C-M diagrams). The granularity is likely to be the
effect of extincting material obscuring the underlying stellar
distribution. Granularity is higher in $J$,
intermediate in $H$ and low/absent in $\ks$ as expected from
the wavelength dependence of extinction. In 5 of the 26 fields where a
single value of reddening for GC stars can account for the observed
colours, no granularity is apparent.

We remark that the presence of extinction towards the GC on far
smaller scales than previously observed, means that an average
extinction correction may in many cases not be valid for photometry of
individual stars in the nuclear bulge.  \citet{schu99} suggested that
smaller-scale structures in the extinction distribution were
responsible for the observed double-peaks in histograms of stellar
number versus $A_V$ in the GC. The findings of this Letter strongly
indicate that extinction may have significantly higher values on
smaller scales than previously measured for the GC. A follow-up paper
by Gosling et al., measures, maps and analyses the extinction at the GC.

A very recent paper by \cite{nish06} has found that IR
extinction varies from sight-line to sight-line towards the GC and
that the universality of IR extinction values are not valid for
the GC region.  Our paper demonstrates that the extinction varies on
even finer scales than they suggest, in a way which correlates with
the granularity of the stellar distribution in the GC.\\

A.J.G.\ thanks PPARC for a studentship. K.M.B.\ thanks the Royal
Society for a University Research Fellowship.  Based on observations
made with the ESO VLT at Paranal 
under programme ID 071.D-0377(A).

\end{document}